# ROOM TEMPERATURE ACCELERATOR STRUCTURES FOR LINEAR COLLIDERS*

R.H. Miller, R.M. Jones, C. Adolphsen, G. Bowden, V. Dolgashev, N. Kroll Z. Li, R. Loewen, C. Ng, C. Pearson, T. Raubenheimer R. Ruth, S. Tantawi, J.W. Wang, SLAC, Menlo Park, CA, U

Abstract

Early tests of short low group velocity and standing wave structures indicated the viability of operating X-band linacs with accelerating gradients in excess of 100 MeV/m. Conventional scaling of traveling wave traveling wave linacs with frequency scales the cell dimensions with $\lambda$. Because Q scales as $\lambda^{1/2}$, the length of the structures scale not linearly but as $\lambda^{3/2}$ in order to preserve the attenuation through each structure. For NLC we chose not to follow this scaling from the SLAC S-band linac to its fourth harmonic at X-band. We wanted to increase the length of the structures to reduce the number of couplers and waveguide drives which can be a significant part of the cost of a microwave linac. Furthermore, scaling the iris size of the disk-loaded structures gave unacceptably high short range dipole wakefields. Consequently, we chose to go up a factor of about 5 in average group velocity and length of the structures, which increases the power fed to each structure by the same factor and decreases the short range dipole wakes by a similar factor. Unfortunately, these longer (1.8 m) structures have not performed nearly as well in high gradient tests as the short structures. We believe we have at least a partial understanding of the reason and will discuss it below. We are now studying two types of short structures with large apertures with moderately good efficiency including: 1) traveling wave structures with the group velocity lowered by going to large phase advance per period with bulges on the iris, 2) $\pi$ mode standing wave structures.

# ROOM TEMPERATURE ACCELERATOR STRUCTURES FOR LINEAR COLLIDERS*

R.H. Miller, R.M. Jones, C. Adolphsen, G. Bowden, V. Dolgashev, N. Kroll Z. Li, R. Loewen, C. Ng, C. Pearson, T. Raubenheimer R. Ruth, S. Tantawi, J.W. Wang, SLAC, Menlo Park, CA, USA

*Abstract*

Early tests of short low group velocity and standing wave structures indicated the viability of operating X-band linacs with accelerating gradients in excess of 100 MeV/m. Conventional scaling of traveling wave traveling wave linacs with frequency scales the cell dimensions with λ. Because Q scales as $\lambda^{1/2}$, the length of the structures scale not linearly but as $\lambda^{3/2}$ in order to preserve the attenuation through each structure. For NLC we chose not to follow this scaling from the SLAC S-band linac to its fourth harmonic at X-band. We wanted to increase the length of the structures to reduce the number of couplers and waveguide drives which can be a significant part of the cost of a microwave linac. Furthermore, scaling the iris size of the disk-loaded structures gave unacceptably high short range dipole wakefields. Consequently, we chose to go up a factor of about 5 in average group velocity and length of the structures, which increases the power fed to each structure by the same factor and decreases the short range dipole wakes by a similar factor. Unfortunately, these longer (1.8 m) structures have not performed nearly as well in high gradient tests as the short structures. We believe we have at least a partial understanding of the reason and will discuss it below. We are now studying two types of short structures with large apertures with moderately good efficiency including: 1) traveling wave structures with the group velocity lowered by going to large phase advance per period with bulges on the iris, 2) π mode standing wave structures.

## 1 HIGH GRADIENT RF BREAKDOWN

The high gradient RF breakdown testing is reported in detail by Adolphsen [1]. There are several interesting and some surprising results that affect the choice of structure design, which we will discuss here. The first is that as suggested by Adolphsen the viable operating gradient appears to vary almost linearly with the inverse of the of the group velocity. A related observation is that the structures process very rapidly with a relatively small number of arcs up to a gradient that also varies slightly less than linearly with the inverse of the group velocity. Above that gradient the arcing rate increases dramatically and progress to higher gradients is very, very slow.

The second and perhaps the most surprising result occurred during the simultaneous testing of a 105cm long structure and a 20cm long structure driven by the same klystron through a 3 dB coupler so that the drive levels and history would be identical. Both structures were designed to be approximately constant gradient, but precisely constant peak surface field on all disks. Both had an initial group velocity of 5% of the velocity of light. The short structure was identical to the first 20 cm of the long structure. One might have expected the 105cm structure to have roughly 5 times as many arcs as the 20cm structure. Instead, the two structures had equal numbers of arcs at all power levels within the statistical variation, except during the very early processing. This is less surprising in view of the fact that the vast majority of the arcs in the long structure occur in the first 20cm.

The third interesting fact emerging from the high gradient testing is observed when a structure has been processed up to some level with a short pulse and the pulse length is increased significantly. The rate of breakdowns increases dramatically with the arcs distributed uniformly in time within the pulse, not concentrated in the added portion of the pulse.

The fourth interesting result occurred in an experiment studying high electric field gradients in rectangular waveguide, Dolgashev [2]. The large dimension of the waveguide had been reduced to lower the group velocity to about 0.18c in order to raise the field strengths that could be reached with available power. The striking observation was that when the pulse length was less than 400ns and the peak surface gradient was 80 MV/m the arcs never degraded the high gradient performance of the waveguide. When the pulse length was more than 500 ns the arcs frequently degraded the high gradient performance. The degradation observed was a higher rate of arcing, or the inability to reach 80MV/m and higher xray levels on pulses where no arcing occurs.

These four observations suggest that it may be important to consider the energy deposited at the site of an arc when an arc occurs for two reasons. First, it may alter the microwave parameters of the structure by causing a tiny, deposited-energy-dependent change in the resonant frequency of the cell in which the arc occurs and thus change the phase advance and the match of the structure. Secondly, there probably is a deposited energy threshold above which the high gradient performance of the structure is degraded. Above this threshold an arc is likely to cause surface damage which causes successive arcs to occur at or in the vicinity of the original site. For many years people have observed that RF processing is not always monotonic; that sometimes an arc causes a setback, lowering the RF power which the device being processed will accept. It has also been realized that it was advantageous to high power process with a short pulse until the desired gradients are reached, and then slowly increase the pulse length. The energy deposited in an arc in a travelling wave structure should vary linearly with the incident power, the pulse length, and the group velocity. It may also depend on other parameters such as the frequency, phase advance per cell and geometric factors such as the gap across which the breakdown occurs. The damage threshold almost surely depends on the physical properties such as melting point of the material from which the structure is made. Despite our ignorance of these issues, it still may be useful to define a damage parameter, D.

*Work supported by the U.S. DOE, Contract DE-AC03-76SF00515.

$$D = PTv_g/c \qquad (1)$$

P is the input power, T is the pulse length, vg is the group velocity and c is the velocity of light. To get an idea of what may reasonable we can look at SLAC (ignoring any frequency and geometry dependence). The highest power operation of the standard SLAC S-band sections with a rectangular pulse (unSLEDded) occurred in the injector where several sections ran routinely with 25 MW 2.5 μs long pulses. The initial group velocity of these constant gradient sections was .02c. With these parameters the damage parameter is 1.2 Joules. It is important to note that only a minuscule fraction of this can be dissipated at the site of the arc. In the high gradient testing the X-band 1.8 meter Damped Detuned Structures (DDS) processed very easily and essentially monotonically up to about 35 MeV/m with a 44 MW, 240ns pulse. The initial group velocity is .12c, giving a damage parameter of 1.3 J, suggesting that this may be a useful parameter. The four observations from the high gradient testing reported above all suggest that for gradients above some damage threshold most of the arcs result from damage caused by previous arcs rather than from the initial existing defects such as inclusions and microscopic points in our copper structures. We have also observed both with the SLAC S-band structures and with the NLC X-band structures that it is possible to have arc damage from which it is impossible to recover. That is that a reasonable amount of high gradient processing cannot recover the gradient at which the structure had been operating. The conservative course of action may be to design to run below the arc damage threshold for which our damage parameter D may (or may not) be a useful indicator. If D were a precise measure of the limit for monotonic processing then the gradient at which monotonic processing ends would scale linearly with the $1/v_g$. We observe a less than linear scaling, but the number of samples is very small. Perhaps $v_g/c$ in the damage parameter needs an experimentally determined exponent which is less than unity.

## 2 TRAVELING WAVE STRUCTURES

The present design for the unloaded gradient for NLC is 72 MeV/m. To achieve this with a damage parameter of the order of 1 Joule will require an initial group velocity in our approximately constant gradient structures of between .03c and .05c. We are presently testing structures at each of these values. The tested structures would not be satisfactory for NLC because the apertures are too small causing excessive dipole wakefields. Z. Li [3] has designed structures for each of these group velocities using 150° phase advance per cell and thicker disks to achieve these lower group velocities with the same average iris diameter as in the $v_g = 0.12c$ structure. The 0.05c structure is 90cm long, while the 0.03c structure is 60 cm long. R.M. Jones [4] is studying detuning these and damping them using either manifold damping or local damping to reduce the long-range dipole wakes by a factor greater than 100. The initial results look quite promising. The wakefield for the 0.05c structure with 10% Gaussian detuning and manifold damping is shown in Fig. 1. It is difficult to go below an initial group velocity of .03c without reducing the average iris diameter, which we don't want to do because of the short-range dipole wakes. We are uncomfortable about going to phase advances larger than 150° per period because of the reduced bandwidth, and thicker disks hurt the shunt impedance.

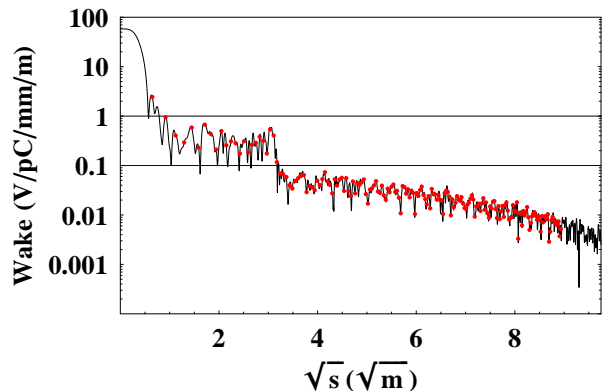

Figure 1: Wakefield for two 90cm long manifold-damped 10% detuned structures with .05c initial $v_g$.

## 3 STANDING WAVE STRUCTURES

There is some argument for designing the NLC structure to operate at unloaded gradients as high as 100 MeV/m, to accommodate an energy upgrade above 1 TeV in the same tunnel. We think this forces us to consider standing wave structures. Standing wave structures have several advantages over travelling for high gradient operation. The first is that for a given gradient the input power required scales roughly as length, and stranding wave structures can be made arbitrarily short without sacrificing efficiency. Secondly, because a standing wave structure is a high Q resonant cavity the reflection coefficient goes very, very close to unity almost instantly when loaded by an arc. In this way a standing wave cavity is more self-protecting than a travelling wave structure. We are studying 20 and 30cm π-mode structures with 15 and 23 cells, respectively. We think this range is a reasonable compromise between tolerances and the costs associated with a large number of short structures. Figure 2 is computer cutaway of our first 15-cell π mode cavity. Two of these are currently in high power testing. We propose to braze 6 to 10 of these short structures together to form a single precision aligned assembly. Each multicell cavity will have an odd number of cells and be driven through the center cell. Center driving relaxes the tolerances by a factor of 4.

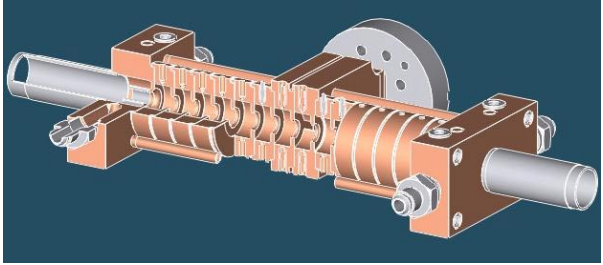

Figure 2: Computer cutaway of 15-cell π-mode cavity.

The standing wave cavities in each half of the assembly will be driven by a single oversize rectangular waveguide with a directional coupler splitting off the appropriate power for each cavity. When this is done right all the power reflected from the cavities goes into dummy loads in the approximation that all the cavities have the same coupling and the same resonant frequency. Thus, in this approximation the assembly looks like a matched load to the RF source. Of course when a cavity arcs it upsets the match, but since each cavity is getting a small fraction of the total power the reflected power goes like the small fraction squared.

## 3.1 Dipole Wakefield Reduction

The same general methods, namely detuning and damping, are available for reducing dipole wakefields in standing wave cavities as were used in the traveling wave structures. However, because the cavities are so short the implementation will be different. Tapering the iris diameters with compensating changes in the cell diameters still appears to be the best way to tune the first dipole band while keeping fundamental π mode frequency constant. By decreasing the thickness of the disks as the iris diameters decrease we find we can reduce the spread in cell to cell coupling of the fundamental. We can get an 8% detuning of the first dipole band with the fundamental mode cell to cell coupling varying from 5.2% at the large iris cells down to 2.2% at small iris end. We intend to vary the iris size and therefore the dipole frequencies monotonically from one end to the other of a 6 or 9 cavity assembly having a total of about 135 cells. Fig. 3 shows the amplitude and phase from an equivalent circuit for each cell of a 15 cell cavity in which the coupling varies from 3.5% to 2.2% as it might in the last cavity at the high dipole frequency end of a 6 cavity assembly. The phase shifts, which are caused by the power flow, can be compensated for by adjusting the length of the cells to keep a velocity of light beam on the crest in each cell. Fig. 4 shows the dispersion curve for each cell (as if it were in a periodic structure). The cell fundamental mode resonant frequencies have been tuned to achieve a flat field, which requires that the π mode frequencies for all the interior cells coincide at about 11.4 GHz. The cells at each end, because they are full cells rather than a half-cells, must be tuned so the π/2 modes are at 11.4 GHz.

Because the cavities are so short, we tried what might be called end damping. We put a low Q (10) dipole mode cavity in the drift tube between each pair of 23 cell cavities and at each end of the full assembly. We have a preliminary simulation of this and the wakefield is presented in Fig. 4. It appears to be promising, but this is certainly neither a full nor optimized design. We have a concept for the lossy dipole cavities but they have not been designed.

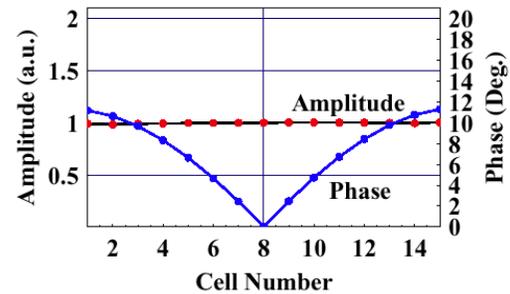

Figure 3: Phase and amplitude of the fundamental mode in the 15 cells of a dipole detuned π mode cavity.

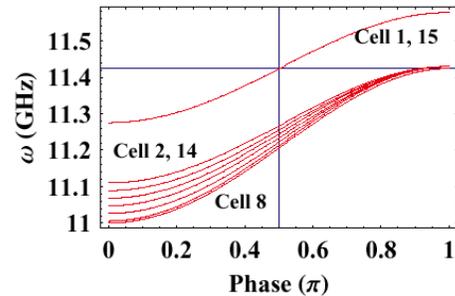

Figure 4: Dispersion curves for the fundamental mode of the 15 cells of a π-mode cavity with dipole mode detuning

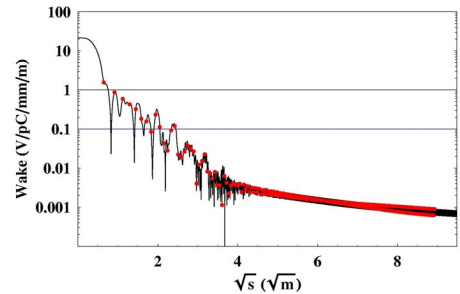

Figure 5: Wakefield for six 23 cell π-mode cavities in a monotonically detuned assembly with a low Q dipole cavity every 23 cells.

## 4 REFERENCES

[1] C. Adolphsen, Paper ROAA003 this conference

[2] V. Dolgashev, Paper FPAH057 this conference
[3] Z. Li, Paper FPAH061 this conference
[4] R.M. Jones, Papers FPAH058 & MPPH068 this conf.